%% file: main.tex
\documentclass[12pt]{article}

\usepackage[margin=25mm]{geometry}
\usepackage{graphicx}
\usepackage{hyperref}
\usepackage{url}
\usepackage{amssymb}
\usepackage{tabularx}
\usepackage{amsmath}
\usepackage{authblk}
\usepackage{mathtools}
\usepackage{amsthm}
\usepackage{blkarray}
\usepackage[numbers]{natbib}
\usepackage{booktabs}
\usepackage{makecell}

\theoremstyle{remark}
\newtheorem*{remark}{Remark}

\input{def}

\begin{document}
\title{A Trajectory-Informed Clustering Approach for Information Borrowing in Basket Trials}
\author[1]{Masahiro Kojima\footnote{Address:1-13-27 Kasuga,Bunkyo-ku,Tokyo 112-8551, Japan. Tel: +81-(0)3-3817-1949 \quad
E-Mail: mkojima263@g.chuo-u.ac.jp}}
\author[2]{Keisuke Hanada}
\author[1]{Atsuya Sato}
\affil[1]{Department of Data Science for Business Innovation, Chuo University}
\affil[2]{Department of Biostatistics, Wakayama Medical University}

\maketitle
\begin{abstract}
Heterogeneity in treatment efficacy is often observed across baskets in basket trials, making appropriate information borrowing challenging. We propose a trajectory-informed clustering framework that groups baskets using transition probabilities derived from longitudinal tumor response patterns, rather than relying solely on a single efficacy endpoint such as the objective response rate (ORR). The number of clusters is selected in a data-driven manner according to the similarity structure among baskets, and baskets assigned to the same cluster are subsequently analyzed using a hierarchical Bayesian model. The proposed framework is intended to improve estimation precision and borrowing efficiency in settings with latent heterogeneity across baskets. In simulation studies, the proposed approach showed improved recovery of heterogeneous cluster structures relative to ORR-only clustering and yielded favorable basket-wise operating characteristics, including improved power with generally acceptable type~I error performance in heterogeneous settings.
\end{abstract}
\par\vspace{4mm}
{\it Keywords: transition probability; basket trial; silhouette method; hierarchical Bayesian model; weighted final-state probability} 

\section{Introduction}
In recent years, there has been growing attention on optimizing dose selection and identifying appropriate target populations in early-phase oncology trials, particularly following initiatives such as the U.S.\ Food and Drug Administration’s (FDA) Project Optimus. Among the master protocol designs (e.g. umbrella, platform, and basket trials) developed to evaluate multiple tumor types, biomarkers, or molecular subtypes in parallel, basket trials have become increasingly important in recent years. In basket trials, efficacy is typically assessed using the objective response rate (ORR) within each basket, which often involves a relatively small sample size of approximately 20 to~50 patients~\cite{Hirakawa2019,FrontiersBasket2023}. Consequently, efficient statistical methods for analyzing data from basket trials are actively being investigated.

A variety of statistical methodologies have been proposed to address the analytical challenges in basket trials, where heterogeneity in treatment effects across tumor types may mask true efficacy signals. Simple independent-basket analyses avoid inappropriate borrowing but may have limited efficiency when baskets are biologically or clinically related \cite{Simon2017CriticalReview,kaizer2019basket}. 
To strike a balance between no borrowing and complete pooling, hierarchical Bayesian models have been widely adopted to allow partial exchangeability of treatment effects and adaptive borrowing of information \cite{Neuenschwander2016,hobbs2018bayesian}. Further refinements, such as the calibrated hierarchical Bayesian (CHB) model \cite{ChuYuan2018CHB} and robust exchangeability approaches based on exchangeable/non-exchangeable components \cite{Neuenschwander2016}, have enabled more flexible control of information borrowing. Related robust testing and model-averaging approaches have also been proposed to mitigate over-borrowing when baskets differ substantially in response \cite{Zhou2021,psioda2021bayesian}. More recently, latent subgroup and partition-based models have been proposed to account for unobserved similarities among baskets. 
Chu and Yuan \cite{Chu2018} introduced the Bayesian Latent Subgroup Design (BLAST), which simultaneously models treatment response and longitudinal biomarkers to detect latent clusters of baskets. Zhou and Ji \cite{Zhou2021} developed the Robust Bayesian Hypothesis Testing (RoBoT) framework, which identifies groups of baskets with similar efficacy profiles. Building on these ideas, Govande and Slate \cite{govande2025using} proposed the Bayesian Partition Model with Covariates (BPMx), which integrates subject-level covariate information to guide latent cluster membership. 
Biomarker-based models offer a promising approach to explaining the sources of heterogeneity; however, in practical basket trials, the relevant and informative biomarkers are often unknown at the time of trial initiation. Furthermore, it is not always clear which covariates are useful for clustering, and including irrelevant variables may obscure the true basket structure. Therefore, developing methods that can identify similarities among baskets using universally collected data without relying on prespecified biomarkers or covariates is of substantial practical importance. Our proposed approach addresses this need by leveraging the trajectories of treatment response, which are typically observed and closely linked to primary efficacy endpoints such as the ORR in basket trials.

In this study, we propose a trajectory-informed clustering approach for basket trials to support data-driven information borrowing across baskets. The proposed method uses transition probabilities derived from longitudinal tumor response trajectories to quantify similarity among baskets, instead of relying only on a single summary endpoint such as the ORR. Baskets identified as similar are subsequently analyzed within a hierarchical Bayesian model so that borrowing is restricted to data-driven borrowing groups. This framework is motivated by small-sample basket settings in which inappropriate full pooling may inflate false positive findings, whereas no borrowing may substantially reduce efficiency. The operating characteristics of the proposed approach are evaluated through simulation studies under several heterogeneous and homogeneous basket configurations.

The remainder of this paper is organized as follows.
Section 2 introduces the calculation of the weighted final-state probabilities of treatment response, which account for varying numbers of assessment time points across patients, followed by descriptions of the clustering procedure and the hierarchical Bayesian model.
Section 3 presents the simulation design and results.
Section 4 discusses the implications and limitations of the proposed method.

\section{Method}
In this section, we describe the proposed two-stage framework for trajectory-informed information borrowing in basket trials. The primary efficacy endpoint considered in the second-stage borrowing model is binary objective response, defined as the occurrence of complete response (CR) or partial response (PR). The ordinal tumor response categories, CR, PR, stable disease (SD), and progressive disease (PD), are used to construct trajectory-derived features for clustering, but the Bayesian borrowing model itself is fitted to the binary ORR endpoint.

\subsection{Transition Probability Matrix and Weighted Final-State Distribution}
We consider a basket trial consisting of $J$ baskets. Let $n_j$ denote the number of patients enrolled in the $j$th basket. For patient $i$ in basket $j$, let
\[
\{X_{ji,1}, \ldots, X_{ji,t_{ji}}\}
\]
denote the observed sequence of tumor response assessments, where $t_{ji}$ is the number of observed assessments and
\[
X_{ji,\ell}\in\mathcal S=\{\mathrm{CR},\mathrm{PR},\mathrm{SD},\mathrm{PD}\}.
\]
An extension that explicitly accounts for not evaluable (NE) cases by augmenting the state space is provided in the Supplementary Material.

We assume that, within each basket, the response process follows a first-order time-homogeneous Markov model over the discrete assessment times:
\[
\Pr(X_{ji,\ell+1}=s \mid X_{ji,\ell}=r,
X_{ji,1},\ldots,X_{ji,\ell-1}, j)
=
\Pr(X_{ji,\ell+1}=s \mid X_{ji,\ell}=r,j)
=
P_j^{rs},
\]
for $r,s\in\mathcal S$. Thus, $P_j^{rs}$ represents the basket-specific one-step transition probability from state $r$ to state $s$, averaged over the scheduled assessment intervals. Under this time-homogeneous assumption, the transition probability after $q$ transitions is represented by the $q$th power of the same transition matrix, $\mathbf P_j^q$.

Let the initial state probability vector in basket $j$ be
\begin{align}
\begin{array}{rcccc}
& CR & PR & SD & PD\\
\bpi_{j0}= &(\pi_{j01}, & \pi_{j02}, & \pi_{j03}, & \pi_{j04}).
\end{array}
\end{align}
Its empirical estimate, $\bpih_{j0}$, is obtained from the distribution of the first observed response assessment in basket $j$.

For a patient with $t_{ji}$ assessments, the number of observed one-step transitions from state $r$ to state $s$ is defined as
\begin{align}
N_{ji}^{rs}(t_{ji})
=
\sum_{\ell=1}^{t_{ji}-1}
I(X_{ji,\ell}=r, X_{ji,\ell+1}=s),
\end{align}
where the sum is zero when $t_{ji}=1$. The transition counts are then aggregated across patients within basket $j$:
\begin{align}
\Nh_{j}^{rs}
=
\sum_{i=1}^{n_j} N_{ji}^{rs}(t_{ji}),
\end{align}
and the total number of observed outgoing transitions from state $r$ is
\begin{align}
\Nh_{j}^{r\cdot}
=
\sum_{s\in\mathcal S}\Nh_{j}^{rs}.
\end{align}
When $\Nh_j^{r\cdot}>0$, the $(r,s)$ element of the estimated one-step transition probability matrix is
\begin{align}
\label{eq:tran_prob}
\Ph_{j}^{rs}
=
\frac{\Nh_{j}^{rs}}{\Nh_{j}^{r\cdot}},
\qquad r,s\in\mathcal S.
\end{align}
This estimator targets the conditional probability
\[
P_j^{rs}
=
\Pr(X_{ji,\ell+1}=s\mid X_{ji,\ell}=r,j),
\]
which is assumed not to depend on the assessment index $\ell$ under the time-homogeneous Markov assumption.

\begin{remark}
In empirical estimation of transition matrices, it may happen that a certain state $r$ in basket $j$ has no observed outgoing transitions, that is, $\Nh_j^{r\cdot}=0$. This means that state $r$ was never observed immediately before a subsequent response assessment in basket $j$; it may occur either because no patient was observed in state $r$ or because observations in state $r$ occurred only at the last available assessment. Since such rows cannot be normalized to valid transition probabilities, we regularize them by replacing the corresponding row of $\bPh_j$ with a uniform distribution:
\[
\Ph_j^{rs}=1/|\mathcal S|,
\qquad s\in\mathcal S.
\]
This normalization is applied to all $s\in\mathcal S$ for the affected row. The replacement is used primarily for numerical stability in sparse small-sample settings. We acknowledge that a uniform replacement may not reflect clinically realistic transition patterns in all applications; therefore, its use should be viewed as a pragmatic approximation when no outgoing transitions are observed from a given state. The potential impact of this assumption is discussed in Section~4.
\end{remark}

In basket trials, the number of response assessments often differs among patients. To account for this variation, we introduce a weighting scheme based on the empirical distribution of the number of assessments in each basket. Let
\[
\mathcal T_j=\{t:\sum_{i=1}^{n_j}I(t_{ji}=t)>0\}
\]
denote the set of observed numbers of assessments in basket $j$. We define
\begin{align}
\wh_{jt}
=
\frac{\sum_{i=1}^{n_j}I(t_{ji}=t)}
{\sum_{u\in\mathcal T_j}\sum_{i=1}^{n_j}I(t_{ji}=u)}
=
\frac{1}{n_j}\sum_{i=1}^{n_j}I(t_{ji}=t),
\qquad t\in\mathcal T_j.
\end{align}

The weighted final-state distribution for basket $j$ is then defined as
\begin{align}
\label{eq:w_f_p}
\bpio_j
=
\sum_{t\in\mathcal T_j}
\wh_{jt}\bpih_{j0}\bPh_j^{t-1}
=
\wh_{j1}\bpih_{j0}
+
\sum_{t\in\mathcal T_j\setminus\{1\}}
\wh_{jt}\bpih_{j0}\bPh_j^{t-1}.
\end{align}
Here, $\bPh_j^{t-1}$ is the $(t-1)$-fold matrix product of the estimated one-step transition matrix in equation~(\ref{eq:tran_prob}); for $t=1$, no transition is applied and the distribution is simply $\bpih_{j0}$.

The vector $\bpio_j$ is interpreted as an estimator of the marginal distribution of the response state at the last observed assessment for a randomly selected patient from basket $j$, standardized to the empirical distribution of the number of assessments in that basket. Specifically, the $s$th component, $\bar\pi_{js}$, estimates
\[
\Pr(X_{ji,T_j}=s\mid j),
\]
where $T_j$ is a random assessment count with empirical distribution $\Pr(T_j=t)=\wh_{jt}$. This interpretation assumes that the number of observed assessments is non-informative conditional on the observed response process. Informative missingness, such as death, progression-related discontinuation, or treatment withdrawal, would require either augmenting the state space with an absorbing dropout or death state or conducting sensitivity analyses.

\subsection{Clustering}
Let $\theta_j$ denote a basket-level efficacy parameter for basket $j$ (e.g., the ORR, disease control rate, or another prespecified endpoint), and let $\hat{\theta}_j$ be its estimator from observed data. For clarity, we illustrate the proposed framework using the ORR, which represents the proportion of patients with either CR or PR, as the primary efficacy endpoint.

We perform clustering using the feature vectors $\y_j = (\bpio_{j}^\top, \hat{\theta}_j)^\top$, where $\bpio_{j}$ is defined in equation~(\ref{eq:w_f_p}) and $\hat{\theta}_j$ represents the observed ORR for basket $j$. The feature vector consists of the four components of the weighted final-state distribution, with the observed ORR additionally included so that the primary efficacy endpoint used for downstream borrowing decisions is explicitly reflected in the clustering stage. The length of $\y_j$ is 5. We define the dissimilarity between two baskets $j$ and $k$ by the \textit{Manhattan distance} (also known as the city-block or $L_1$ distance),
\begin{align}
d(\y_j, \y_k) = \sum_{m=1}^5 |y_{jm} - y_{km}|.
\end{align}
The distance has been shown to be more robust to outliers and less sensitive to extreme deviations than the Euclidean distance \cite{Jain1988,KaufmanRousseeuw1990,Everitt2011}. We note, however, that this representation treats the response categories as nominal rather than fully ordinal; this choice was made to preserve a simple and practically interpretable clustering rule in small-sample settings, and its implications are revisited in Section~4. All five components of $\y_j$ are probabilities on $[0,1]$; therefore, no additional scaling is applied prior to computing the Manhattan distance.

We determine the number of clusters using the Silhouette method~\cite{Rousseeuw1987}. For each observation $j$, let $C(j)$ denote the cluster to which it belongs. The average dissimilarity between $j$ and all other observations within the same cluster is defined as
\begin{align}
a_j
= \frac{1}{|C(j)| - 1}
\sum_{\substack{k \in C(j) \\ k \ne j}}
d(\y_j, \y_k),
\end{align}
and the minimum average dissimilarity between $j$ and all points in other clusters $C' \neq C(j)$ is given by
\begin{align}
b_j
= \min_{C' \neq C(j)}
\left\{
\frac{1}{|C'|}
\sum_{k \in C'} d(\y_j, \y_k)
\right\}.
\end{align}
The Silhouette value for observation $j$ is then defined as
\begin{align}
s_j = \frac{b_j - a_j}{\max(a_j, b_j)}, \qquad -1 \le s_j \le 1.
\end{align}
The average Silhouette index
\begin{align}
\bar{s} = \frac{1}{J} \sum_{k=1}^{J} s_k
\end{align}
quantifies the overall cluster separation. To avoid instability due to extremely small clusters, the number of clusters $u$ is first selected by maximizing the average Silhouette index $\bar{s}$ over $u \in \{2, \ldots, J-1\}$, subject to a minimum cluster size of 2. The Silhouette index requires well-defined within-cluster dissimilarities, i.e., $|C(j)| \ge 2$ for every cluster so that $a_j$ is finite. Therefore, the all-singleton partition ($u = J$) is excluded from consideration. In cases where singleton clusters appear, the overall clustering performance is evaluated based on the Silhouette values of non-singleton clusters, while singleton clusters are retained in the clustering result. Following \cite{KaufmanRousseeuw1990}, when the resulting maximum average Silhouette index is small ($\bar{s} \le 0.25$), the clustering structure is regarded as weak or unreliable, and the analysis defaults to a single-cluster borrowing structure. This rule is adopted as a pragmatic safeguard against unstable over-partitioning in very small samples, although its potential limitations under homogeneous and heterogeneous settings are discussed in Section~4.

\subsection{Hierarchical Bayesian Modeling within Clusters}
After clustering is performed based on the feature vectors constructed from the weighted final-state probabilities and the observed ORR, baskets assigned to the same estimated cluster are analyzed jointly using a hierarchical Bayesian model. Importantly, information borrowing is conducted only for the primary efficacy endpoint, namely the ORR, and not for the full tumor response categories (CR, PR, SD, PD), which are used exclusively for clustering. This separation between the clustering stage and the borrowing stage is intentional. In basket trials, efficacy decisions are typically based on a prespecified primary endpoint such as the ORR, whereas response trajectories provide auxiliary information about similarity across baskets. Accordingly, trajectory information is used to construct data-driven borrowing groups, while Bayesian borrowing itself is performed only for the primary efficacy endpoint.

Let
\[
Y_j=\sum_{i=1}^{n_j} I(\text{patient } i \text{ in basket } j \text{ achieves CR or PR})
\]
denote the number of objective responses in basket $j$. Conditional on the estimated cluster partition $\hat{\mathcal C}$, we model
\[
Y_j \mid p_j \sim \mathrm{Binomial}(n_j,p_j),
\qquad
\mathrm{logit}(p_j)=\eta_j.
\]
For baskets assigned to the same estimated cluster $c$, we assume
\[
\eta_j \mid \mu_c,\tau_c \sim \mathcal N(\mu_c,\tau_c^{-1}),
\qquad j\in \hat{\mathcal C}_c,
\]
where $\mu_c$ represents the cluster-specific mean log-odds of response and $\tau_c$ is the precision parameter governing the degree of information sharing within cluster $c$. The hyperpriors are specified as
\[
\mu_c \sim \mathcal N(\mu_0,\sigma_0^2),
\qquad
\tau_c \sim \mathrm{Gamma}(\alpha,\beta),
\]
with the gamma distribution parameterized in terms of shape and rate.

This hierarchical structure induces partial exchangeability among baskets within the same estimated cluster, allowing adaptive borrowing of information while preserving heterogeneity across clusters. When all baskets are assigned to a single cluster, the model reduces to a fully exchangeable hierarchical Bayesian model. Conversely, when baskets are separated into different clusters, borrowing is restricted within each estimated cluster, and no information is borrowed across clusters.

Because the cluster partition is estimated before fitting the hierarchical model, the posterior distributions obtained in the second stage should be interpreted as conditional posterior summaries given the estimated partition, rather than as fully Bayesian posterior distributions marginalizing over cluster uncertainty. The proposed procedure is therefore best viewed as a two-stage, data-adaptive borrowing algorithm. Its inferential properties are assessed through repeated simulation, which captures both the variability of the clustering step and the subsequent conditional Bayesian analysis.

\section{Simulation}
\subsection{Simulation configuration}
We considered five baskets in the simulation and denoted the true transition probability matrix of the $j$th basket by $\P_j$. Three scenarios were examined. In \textbf{Scenario 1}, all baskets shared the same true transition probability matrix. In \textbf{Scenario 2}, three of the baskets shared one matrix, whereas the remaining two baskets shared another matrix identical between them. In \textbf{Scenario 3}, two baskets shared one matrix, another two baskets shared a second matrix, and the fifth basket had a distinct matrix. The specific transition probability matrices and initial distributions were as follows.

\vspace{1em}
\noindent\textbf{[Scenario 1]}
\begin{align}
\boldsymbol\pi_{j0}&=(0.05,\,0.10,\,0.35,\,0.50), \quad \text{for all } j=1,\dots,5,\\
\mathbf P_j &=
\begin{blockarray}{ccccc}
& \text{CR} & \text{PR} & \text{SD} & \text{PD} \\
\begin{block}{c[cccc]}
\text{CR} & 0.60 & 0.00 & 0.00 & 0.40\\
\text{PR} & 0.10 & 0.40 & 0.10 & 0.40\\
\text{SD} & 0.05 & 0.20 & 0.40 & 0.35\\
\text{PD} & 0.00 & 0.05 & 0.35 & 0.60\\
\end{block}
\end{blockarray}
\end{align}

\noindent\textbf{[Scenario 2]}
\begin{align}
\boldsymbol\pi_{j0}&=(0.05,\,0.10,\,0.35,\,0.50), \quad j=1,2,3,\\
\mathbf P_j &=
\begin{blockarray}{ccccc}
& \text{CR} & \text{PR} & \text{SD} & \text{PD} \\
\begin{block}{c[cccc]}
\text{CR} & 0.60 & 0.00 & 0.00 & 0.40\\
\text{PR} & 0.10 & 0.40 & 0.10 & 0.40\\
\text{SD} & 0.05 & 0.20 & 0.40 & 0.35\\
\text{PD} & 0.00 & 0.05 & 0.35 & 0.60\\
\end{block}
\end{blockarray}
\end{align}

\begin{align}
\boldsymbol\pi_{j0}&=(0.075,\,0.150,\,0.425,\,0.350), \quad j=4,5,\\
\mathbf P_j &=
\begin{blockarray}{ccccc}
& \text{CR} & \text{PR} & \text{SD} & \text{PD} \\
\begin{block}{c[cccc]}
\text{CR} & 0.675 & 0.000 & 0.000 & 0.325\\
\text{PR} & 0.150 & 0.475 & 0.100 & 0.275\\
\text{SD} & 0.075 & 0.250 & 0.400 & 0.275\\
\text{PD} & 0.025 & 0.100 & 0.325 & 0.550\\
\end{block}
\end{blockarray}
\end{align}

\noindent\textbf{[Scenario 3]}
\begin{align}
\boldsymbol\pi_{j0}&=(0.05,\,0.10,\,0.35,\,0.50), \quad j=1,2,\\
\mathbf P_j &=
\begin{blockarray}{ccccc}
& \text{CR} & \text{PR} & \text{SD} & \text{PD} \\
\begin{block}{c[cccc]}
\text{CR} & 0.60 & 0.00 & 0.00 & 0.40\\
\text{PR} & 0.10 & 0.40 & 0.10 & 0.40\\
\text{SD} & 0.05 & 0.20 & 0.40 & 0.35\\
\text{PD} & 0.00 & 0.05 & 0.35 & 0.60\\
\end{block}
\end{blockarray}
\end{align}

\begin{align}
\boldsymbol\pi_{j0}&=(0.075,\,0.150,\,0.425,\,0.350), \quad j=3,4,\\
\mathbf P_j &=
\begin{blockarray}{ccccc}
& \text{CR} & \text{PR} & \text{SD} & \text{PD} \\
\begin{block}{c[cccc]}
\text{CR} & 0.675 & 0.000 & 0.000 & 0.325\\
\text{PR} & 0.150 & 0.475 & 0.100 & 0.275\\
\text{SD} & 0.075 & 0.250 & 0.400 & 0.275\\
\text{PD} & 0.025 & 0.100 & 0.325 & 0.550\\
\end{block}
\end{blockarray}
\end{align}

\begin{align}
\boldsymbol\pi_{50}&=(0.10,\,0.20,\,0.50,\,0.20),\\
\mathbf P_{5}&=
\begin{blockarray}{ccccc}
& \text{CR} & \text{PR} & \text{SD} & \text{PD} \\
\begin{block}{c[cccc]}
\text{CR} & 0.75 & 0.00 & 0.00 & 0.25\\
\text{PR} & 0.20 & 0.55 & 0.10 & 0.15\\
\text{SD} & 0.10 & 0.30 & 0.40 & 0.20\\
\text{PD} & 0.05 & 0.15 & 0.30 & 0.50\\
\end{block}
\end{blockarray}
\end{align}

The number of observation time points per patient was sampled from $\{1,2,\dots,10\}$ with probabilities $(0.03,\,0.05,\,0.25,\,0.25,\,0.20,\,0.10,\,0.05,\,0.03,\,0.02,\,0.02)$. The ORR was defined as the probability of ever reaching CR or PR during the observation period. Each simulation setting was evaluated using $R=10{,}000$ simulation replicates.

\vspace{0.8em}
\noindent
As comparators, we also evaluated four alternative models: 
\begin{enumerate}
  \item a clustering approach based solely on the ORR values (\textbf{ORR-only}),
  \item a hierarchical Bayesian model assuming a single common cluster across all baskets, corresponding to complete exchangeability (\textbf{One cluster}),
  \item a model that does not perform clustering, applying independent $\mathrm{Beta}(1,1)$ priors to each basket (\textbf{No cluster}), and
  \item an exchangeable/non-exchangeable Bayesian model (\textbf{EXNEX}).
\end{enumerate}
For the ORR-only approach, clustering was performed using the observed ORR alone and the same silhouette-based cluster selection procedure, following the general strategy of data-driven homogeneous subgroup identification described by Krajewska and Rauch~\cite{Krajewska2021}.

As a supplementary comparator, we considered an EXNEX model following the exchangeable/non-exchangeable framework of Neuenschwander et al.~\citep{Neuenschwander2016}. For each basket, the log-odds of response was modeled as arising from a mixture of an exchangeable component and a basket-specific non-exchangeable component. The exchangeable component was modeled using a logistic-normal hierarchical Bayesian model with
\[
\theta_j \mid \mu,\tau \sim \mathcal{N}(\mu,\tau^{-1}), \qquad
\mu \sim \mathcal{N}(0,1), \qquad
\tau \sim \mathrm{Gamma}(2,1),
\]
whereas the non-exchangeable component was modeled independently with $\theta_j \sim \mathcal{N}(0,1)$.

The prior probability of exchangeability for each basket was set to 0.5.

The hierarchical Bayesian model with a single common cluster represents the special case of full exchangeability among baskets, allowing complete borrowing of information across strata. This model serves as a reference for evaluating the potential gain or loss in performance attributable to the adaptive clustering in the proposed method. For the Proposed, ORR-only, and One-cluster methods, basket-specific ORRs were modeled using a logistic-normal hierarchical Bayesian model. The prior distributions were $\mu \sim \mathcal{N}(0,1)$ and $\tau \sim \mathrm{Gamma}(2,1)$.

The efficacy threshold $c_0$ for declaring an active basket was set to $0.467$ for the ORR instantiation. 
This value corresponds to the approximate true ORR estimated by Monte Carlo simulation with $n_{j}=10{,}000$ 
using $20{,}000$ replications under the reference transition matrix of Scenario~1. 
The threshold was chosen to evaluate the type~I error rate of the proposed method under the null hypothesis 
that the true efficacy equals this reference value. In practice, basket trial designs often declare efficacy based on the posterior probability that the ORR exceeds a prespecified null response rate~\citep{Pohl2021}. However, for the purpose of method comparison in simulation 
studies, operating characteristics such as the type~I error rate and statistical power provide a more 
transparent and standardized performance metric across competing methods. Accordingly, in this study, 
the type~I error rate and power were evaluated based on the probability of declaring a basket active, 
where power was defined as the probability that the lower bound of the 90\% credible interval for the ORR 
exceeds the efficacy threshold. This choice facilitates direct comparison of error control and efficiency 
across methods, while remaining consistent with commonly used Bayesian decision criteria. To design the simulation settings, we submitted several applications to access clinical trial data. 
However, because basket trials are a relatively new design and are actively being used in ongoing drug 
development, access to actual trial data was not granted. Therefore, we conducted supplementary simulations 
to verify the robustness of the main simulation results.

\subsection{Simulation results}
The results of clustering based on the Silhouette method are summarized in Table~\ref{tab:clustering_results}. 
When the true basket structure consisted of a single cluster, the silhouette-based procedure frequently selected more than one cluster, indicating a tendency toward over-partitioning in homogeneous settings. In contrast, under Scenarios~2 and~3, where latent heterogeneity was present, the proposed method more frequently recovered the underlying grouping structure than the ORR-only comparator, although recovery was imperfect, particularly in the three-cluster setting.

\begin{figure}[!ht]
  \centering
  \includegraphics[width=0.85\textwidth]{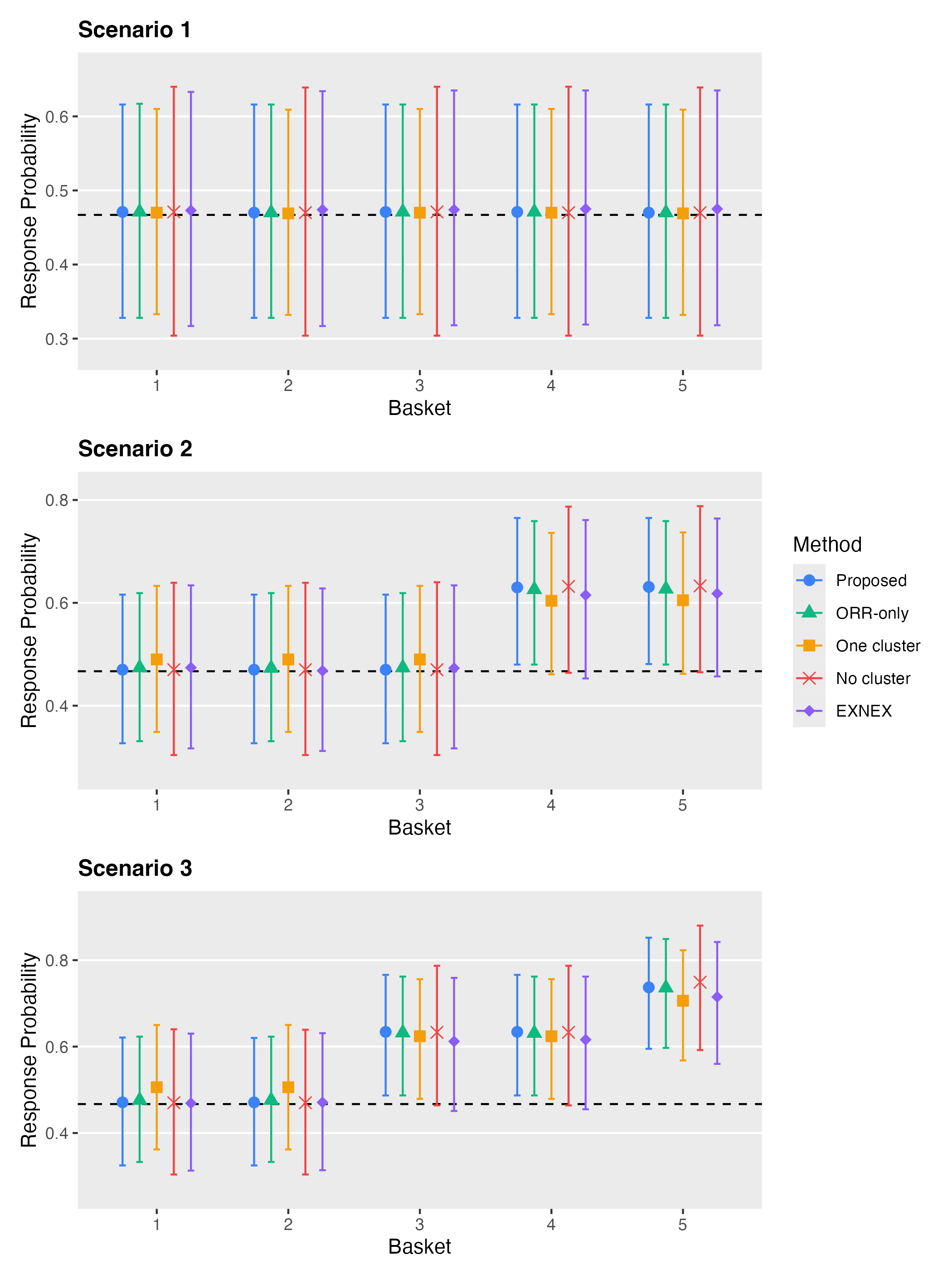}
  \caption{Average posterior mean and 90\% credible intervals across baskets, based on 10,000 simulation replicates, $n_j=20$}
  \label{fig:scenarios_nj20}
  \par\vspace{0.3em}\footnotesize\textit{Note.} The horizontal dashed line indicates the decision threshold $c_0=0.4670$. Posterior summaries are averaged over 10,000 simulation replicates.
\end{figure}

\begin{figure}[!ht]
  \centering
  \includegraphics[width=0.85\textwidth]{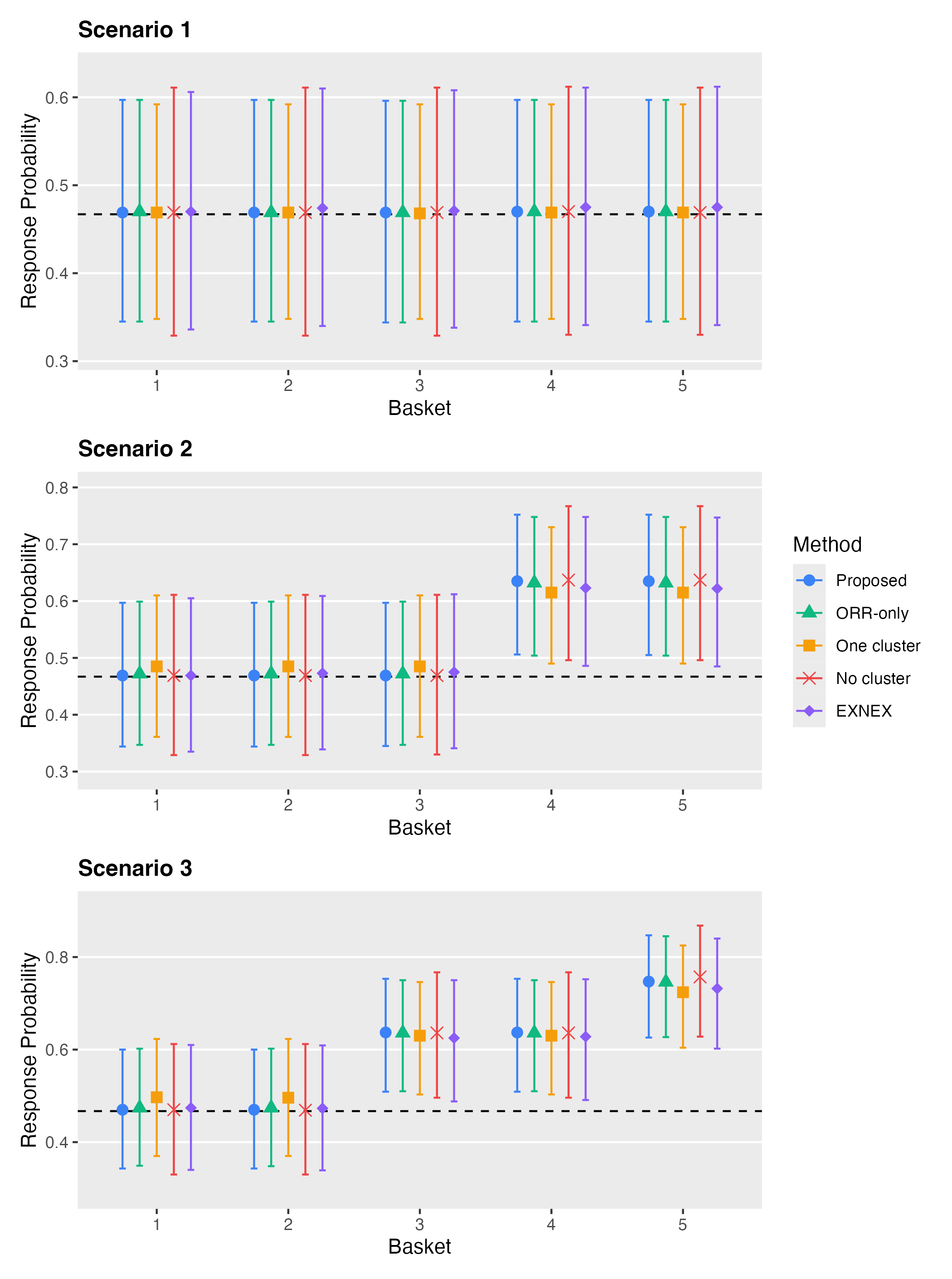}
  \caption{Average posterior mean and 90\% credible intervals across baskets, based on 10,000 simulation replicates, $n_j=30$}
  \label{fig:scenarios_nj30}
  \par\vspace{0.3em}\footnotesize\textit{Note.} The horizontal dashed line indicates the decision threshold $c_0=0.4670$. Posterior summaries are averaged over 10,000 simulation replicates.
\end{figure}

\begin{figure}[!ht]
  \centering
  \includegraphics[width=0.85\textwidth]{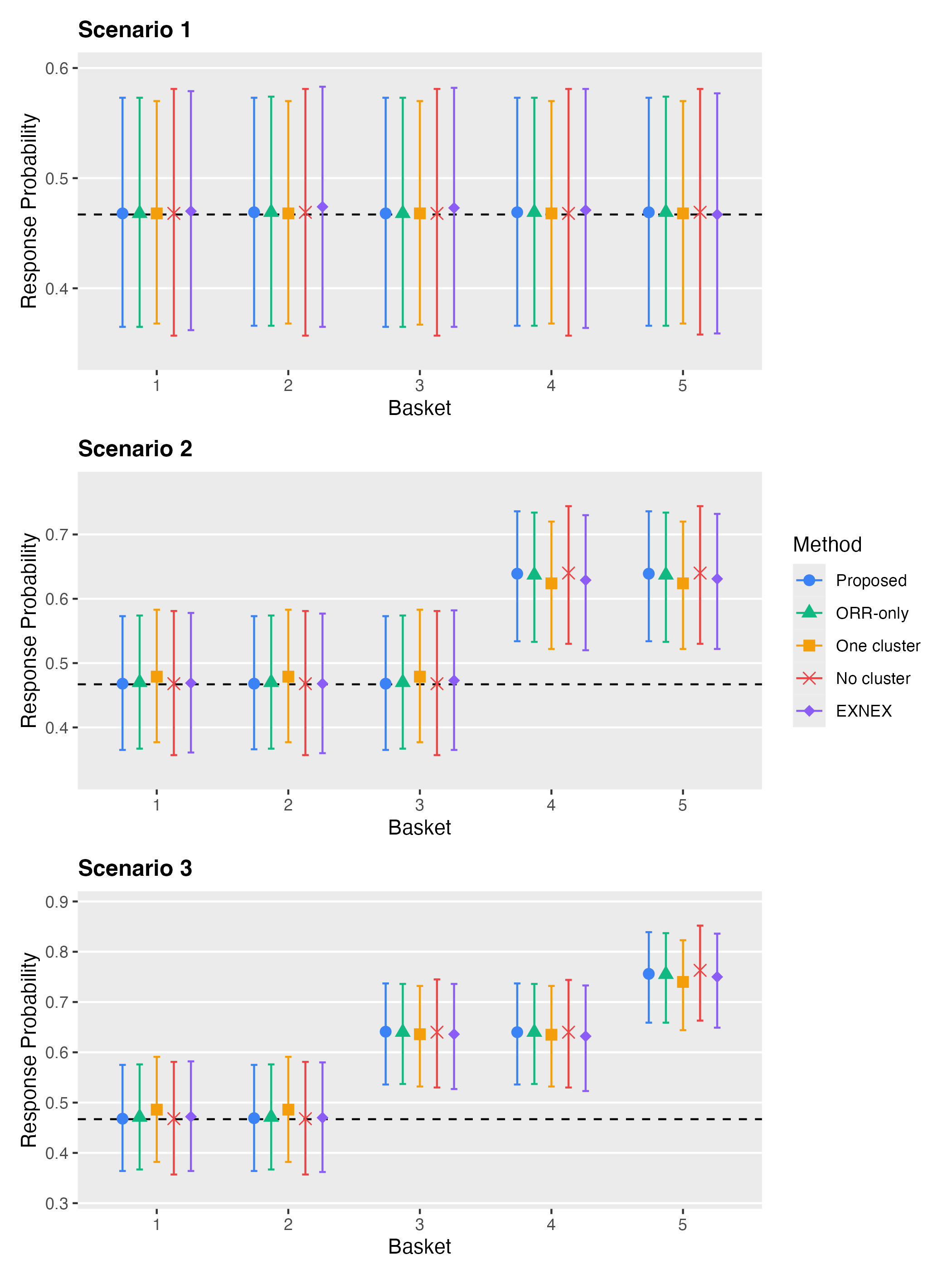}
  \caption{Average posterior mean and 90\% credible intervals across baskets, based on 10,000 simulation replicates, $n_j=50$}
  \label{fig:scenarios_nj50}
  \par\vspace{0.3em}\footnotesize\textit{Note.} The horizontal dashed line indicates the decision threshold $c_0=0.4670$. Posterior summaries are averaged over 10,000 simulation replicates.
\end{figure}

The probabilities of exceeding the efficacy threshold are shown in Table~\ref{tab:cilprob_20}, \ref{tab:cilprob_30}, and \ref{tab:cilprob_50}. 
The posterior means and the average 90\% credible intervals obtained from the simulation are presented in Figures~\ref{fig:scenarios_nj20}, \ref{fig:scenarios_nj30}, and~\ref{fig:scenarios_nj50}. The EXNEX comparator showed behavior closer to that of the No-cluster approach, indicating comparatively limited borrowing in these simulation settings.

More detailed simulation results are provided in the Supplementary Material. 
The design and results of the additional simulations conducted with a lower true efficacy probability are also described in the Supplementary Material.

\begin{table}[!ht]
\centering
\caption{Proportion of clustering results (\%), based on 10,000 simulation replicates}
\label{tab:clustering_results}
\begin{tabular}{l|cc|cc|cc}
\toprule
 & \multicolumn{2}{c}{Scenario 1} & \multicolumn{2}{c}{Scenario 2} & \multicolumn{2}{c}{Scenario 3} \\
\cmidrule(lr){2-3} \cmidrule(lr){4-5} \cmidrule(lr){6-7}
\# clusters & Proposed & ORR-only & Proposed & ORR-only & Proposed & ORR-only \\
\midrule
\multicolumn{7}{l}{\textit{$n_j=20$}}\\
1 & \textbf{0.02} & \textbf{0.05} & 0.00 & 0.05 & 0.17 & 0.08 \\
2 & 77.00 & 74.59 & \makecell{\textbf{94.79}\\ \textbf{(92.89)}} & \makecell{\textbf{78.49}\\ \textbf{(21.57)}} & 48.69 & 78.88 \\
3 & 21.56 & 23.31 & 5.14 & 20.02 & \makecell{\textbf{50.59}\\ \textbf{(50.41)}} & \makecell{\textbf{19.66}\\ \textbf{(5.15)}} \\
4 & 1.42 & 2.05 & 0.06 & 1.44 & 0.54 & 1.38 \\\hline
\multicolumn{7}{l}{\textit{$n_j=30$}}\\
1 & \textbf{0.05} & \textbf{0.06} & 0.00 & 0.10 & 0.03 & 0.11 \\
2 & 79.53 & 76.86 & \makecell{\textbf{97.89}\\ \textbf{(97.47)}} & \makecell{\textbf{80.62}\\ \textbf{(29.90)}} & 42.35 & 80.22 \\
3 & 19.31 & 21.50 & 2.10 & 18.20 & \makecell{\textbf{57.42}\\ \textbf{(57.38)}} & \makecell{\textbf{18.74}\\ \textbf{(7.36)}} \\
4 & 1.12 & 1.57 & 0.01 & 1.08 & 0.21 & 0.93 \\\hline
\multicolumn{7}{l}{\textit{$n_j=50$}}\\
1 & \textbf{0.06} & \textbf{0.02} & 0.00 & 0.03 & 0.00 & 0.03 \\
2 & 81.29 & 78.42 & \makecell{\textbf{99.56}\\ \textbf{(99.53)}} & \makecell{\textbf{84.39}\\ \textbf{(44.47)}} & 34.40 & 80.69 \\
3 & 17.79 & 20.31 & 0.44 & 14.90 & \makecell{\textbf{65.57}\\ \textbf{(65.57)}} & \makecell{\textbf{18.47}\\ \textbf{(11.34)}} \\
4 & 0.86 & 1.25 & 0.00 & 0.69 & 0.04 & 0.81 \\
\bottomrule
\end{tabular}
\vspace{1ex}
\\ \footnotesize{\textbf{Proposed}: silhouette-based clustering on $(\bar{\boldsymbol{\pi}}_j,\hat{\theta}_j)$;
\textbf{ORR-only}: clustering using only ORR. Bold values indicate the proportion correctly identifying the true number of clusters. Values in parentheses represent the proportion correctly capturing the true cluster structure.}
\end{table}

\begin{table}[!ht]
\setlength{\tabcolsep}{4pt}
\centering
\caption{Probability that the 90\% credible interval lower bound exceeds the threshold (\%), based on 10,000 simulation replicates, $n_j=20$}
\label{tab:cilprob_20}
\begin{tabular}{|l|c|c|c|}
\hline
 & Scenario 1 & Scenario 2 & Scenario 3 \\
\hline
\multicolumn{1}{|l|}{Proposed} & & & \\
1 & 7.3 & 4.6 & 5.0 \\
2 & 7.2 & 4.6 & 5.1 \\
3 & 7.3 & 4.7 & 59.3 \\
4 & 7.3 & 56.5 & 59.3 \\
5 & 7.2 & 56.8 & 93.6 \\
\hline
\multicolumn{1}{|l|}{ORR-only} & & & \\
1 & 7.3 & 8.1 & 8.9 \\
2 & 7.2 & 8.1 & 8.9 \\
3 & 7.3 & 8.2 & 58.7 \\
4 & 7.3 & 57.5 & 58.9 \\
5 & 7.2 & 57.7 & 93.7 \\
\hline
\multicolumn{1}{|l|}{One cluster} & & & \\
1 & 3.5 & 6.4 & 10.0 \\
2 & 3.5 & 6.4 & 10.0 \\
3 & 3.5 & 6.5 & 56.3 \\
4 & 3.6 & 46.3 & 56.4 \\
5 & 3.5 & 46.7 & 91.0 \\
\hline
\multicolumn{1}{|l|}{No cluster} & & & \\
1 & 3.1 & 3.0 & 3.0 \\
2 & 3.1 & 3.1 & 3.1 \\
3 & 3.1 & 3.2 & 40.0 \\
4 & 3.2 & 39.9 & 40.1 \\
5 & 3.1 & 40.4 & 85.6 \\
\hline
\multicolumn{1}{|l|}{EXNEX} & & & \\
1 & 3.1 & 2.9 & 3.5 \\
2 & 2.8 & 3.0 & 2.5 \\
3 & 2.6 & 2.5 & 39.3 \\
4 & 3.6 & 39.8 & 41.1 \\
5 & 2.4 & 41.1 & 84.5 \\
\hline
\end{tabular}
\vspace{1ex}\\\footnotesize{\textit{Note.} Entries show $\Pr(\text{90\% CI lower bound}>\text{threshold})$ in percent for each basket and scenario; threshold $=0.4670$.
\textbf{Proposed}: silhouette-based clustering on $(\bar{\boldsymbol{\pi}}_j,\hat{\theta}_j)$;
\textbf{ORR-only}: clustering using only ORR;
\textbf{One cluster}: hierarchical Bayes with complete exchangeability (single cluster, full borrowing);
\textbf{No cluster}: independent per-basket analyses with Beta$(1,1)$ (no borrowing);
\textbf{EXNEX}: exchangeable/non-exchangeable Bayesian model.}
\end{table}

\begin{table}[!ht] 
\setlength{\tabcolsep}{4pt}
\centering
\caption{Probability that the 90\% credible interval lower bound exceeds the threshold (\%), based on 10,000 simulation replicates, $n_j=30$}
\label{tab:cilprob_30}
\begin{tabular}{|l|c|c|c|}
\hline
 & Scenario 1 & Scenario 2 & Scenario 3 \\
\hline
\multicolumn{1}{|l|}{Proposed} & & & \\
1 & 5.5 & 4.8 & 5.1 \\
2 & 5.5 & 4.7 & 5.1 \\
3 & 5.5 & 4.7 & 68.9 \\
4 & 5.4 & 68.2 & 68.8 \\
5 & 5.5 & 68.1 & 98.2 \\
\hline
\multicolumn{1}{|l|}{ORR-only} & & & \\
1 & 5.5 & 7.5 & 7.3 \\
2 & 5.5 & 7.4 & 7.3 \\
3 & 5.5 & 7.4 & 68.2 \\
4 & 5.3 & 67.3 & 68.2 \\
5 & 5.4 & 67.1 & 98.3 \\
\hline
\multicolumn{1}{|l|}{One cluster} & & & \\
1 & 3.8 & 6.4 & 9.4 \\
2 & 3.9 & 6.3 & 9.3 \\
3 & 3.9 & 6.3 & 68.9 \\
4 & 3.7 & 62.6 & 68.8 \\
5 & 3.8 & 62.4 & 98.0 \\
\hline
\multicolumn{1}{|l|}{No cluster} & & & \\
1 & 5.1 & 5.1 & 5.1 \\
2 & 5.1 & 5.0 & 5.1 \\
3 & 5.1 & 5.0 & 63.4 \\
4 & 5.0 & 63.7 & 63.5 \\
5 & 5.1 & 63.5 & 97.6 \\
\hline
\multicolumn{1}{|l|}{EXNEX} & & & \\
1 & 4.1 & 5.2 & 5.4 \\
2 & 5.8 & 6.0 & 4.59 \\
3 & 4.7 & 5.1 & 63.2 \\
4 & 5.1 & 63.5 & 65.4 \\
5 & 4.9 & 62.7 & 98.3 \\
\hline
\end{tabular}
\vspace{1ex}\\\footnotesize{\textit{Note.} Entries show $\Pr(\text{90\% CI lower bound}>\text{threshold})$ in percent for each basket and scenario; threshold $=0.4670$.
\textbf{Proposed}: silhouette-based clustering on $(\bar{\boldsymbol{\pi}}_j,\hat{\theta}_j)$;
\textbf{ORR-only}: clustering using only ORR;
\textbf{One cluster}: hierarchical Bayes with complete exchangeability (single cluster, full borrowing);
\textbf{No cluster}: independent per-basket analyses with Beta$(1,1)$ (no borrowing);
\textbf{EXNEX}: exchangeable/non-exchangeable Bayesian model.}
\end{table}

\begin{table}[!ht]   
\setlength{\tabcolsep}{4pt}
\centering
\caption{Probability that the 90\% credible interval lower bound exceeds the threshold (\%), based on 10,000 simulation replicates, $n_j=50$}
\label{tab:cilprob_50}
\begin{tabular}{|l|c|c|c|}
\hline
 & Scenario 1 & Scenario 2 & Scenario 3 \\
\hline
\multicolumn{1}{|l|}{Proposed} & & & \\
1 & 5.6 & 4.4 & 4.6 \\
2 & 5.8 & 4.5 & 4.6 \\
3 & 5.7 & 4.4 & 86.5 \\
4 & 5.6 & 86.2 & 86.4 \\
5 & 5.7 & 86.1 & 99.9 \\
\hline
\multicolumn{1}{|l|}{ORR-only} & & & \\
1 & 5.6 & 6.5 & 6.8 \\
2 & 5.8 & 6.6 & 6.9 \\
3 & 5.7 & 6.6 & 84.1 \\
4 & 5.6 & 85.0 & 83.9 \\
5 & 5.7 & 84.9 & 99.9 \\
\hline
\multicolumn{1}{|l|}{One cluster} & & & \\
1 & 4.1 & 6.3 & 7.9 \\
2 & 4.2 & 6.5 & 8.0 \\
3 & 4.1 & 6.4 & 86.3 \\
4 & 4.0 & 81.9 & 86.0 \\
5 & 4.1 & 81.8 & 99.9 \\
\hline
\multicolumn{1}{|l|}{No cluster} & & & \\
1 & 4.1 & 4.0 & 4.0 \\
2 & 4.2 & 4.1 & 4.1 \\
3 & 4.1 & 4.0 & 79.6 \\
4 & 4.0 & 79.7 & 79.5 \\
5 & 4.1 & 79.6 & 99.8 \\
\hline
\multicolumn{1}{|l|}{EXNEX} & & & \\
1 & 3.6 & 3.5 & 3.8 \\
2 & 4.9 & 3.8 & 4.3 \\
3 & 4.0 & 4.8 & 80.8 \\
4 & 4.8 & 79.4 & 77.9 \\
5 & 3.8 & 79.2 & 99.9 \\
\hline
\end{tabular}
\vspace{1ex}\\\footnotesize{\textit{Note.} Entries show $\Pr(\text{90\% CI lower bound}>\text{threshold})$ in percent for each basket and scenario; threshold $=0.4670$.
\textbf{Proposed}: silhouette-based clustering on $(\bar{\boldsymbol{\pi}}_j,\hat{\theta}_j)$;
\textbf{ORR-only}: clustering using only ORR;
\textbf{One cluster}: hierarchical Bayes with complete exchangeability (single cluster, full borrowing);
\textbf{No cluster}: independent per-basket analyses with Beta$(1,1)$ (no borrowing);
\textbf{EXNEX}: exchangeable/non-exchangeable Bayesian model.}
\end{table}

\section{Discussion}
In this study, we proposed a novel clustering approach for basket trials that incorporates the trajectories of treatment responses, which have not previously been utilized for clustering in this context. Because treatment response trajectories serve as the foundation for calculating the objective response rate (ORR)—which is often the primary efficacy endpoint in basket trials—they are naturally correlated with the ORR. Previous studies have proposed clustering methods based solely on the ORR; however, relying on this single binary proportion may result in unstable clustering performance, particularly when sample sizes are small or when response probabilities are close to the decision threshold. Other approaches utilizing biomarkers or covariates have also been proposed, but in practice, effective biomarkers may not be known in advance, and it is often unclear which covariates are important for clustering. 
The proposed method overcomes these limitations by leveraging the treatment response trajectories, which are routinely collected in basket trials. Through simulation studies, we found that when two or more latent clusters were present, integrating treatment response trajectories into the clustering framework improved recovery of the underlying grouping structure relative to ORR-only clustering and yielded favorable basket-wise operating characteristics in the heterogeneous settings considered. Furthermore, when estimating the final-state distribution of treatment responses, we accounted for variation in the number of response assessments across patients by introducing a schedule-weighted estimation based on observation frequency. This weighting strategy provided more reliable estimates of the final-state probabilities for each basket, thereby stabilizing the subsequent clustering process.

We also explored alternative clustering approaches such as Partitioning Around Medoids (PAM)~\cite{KaufmanRousseeuw1990} and Bayesian mixture models, including the Dirichlet Process Mixture Model (DPMM)~\cite{Li2019DPMM}. However, with very small sample sizes, which are common in basket trials, these methods often yielded unstable or inconsistent cluster assignments. In addition, we considered applying a generalized mixture model to cluster based on the transitions of treatment response. However, we ultimately decided not to adopt this approach because we aimed to perform clustering that simultaneously incorporates both treatment response transitions and the overall efficacy measure, such as the ORR.

In the simulation studies, under Scenario~1—where only a single cluster existed—the proposed silhouette-based clustering procedure frequently selected more than one cluster. This indicates that the current implementation is prone to over-partitioning in homogeneous settings, and this behavior was associated with a slight inflation of the type~I error rate beyond the one-sided nominal 5\% level. In contrast, in Scenarios~2 and~3, where multiple cluster structures were present, the proposed method more frequently recovered the intended grouping structure than simpler comparators, while yielding type~I error rates that were generally close to the one-sided nominal level. When clustering was performed using only the ORR, as shown in Table~\ref{tab:clustering_results}, this approach failed to adequately capture the true cluster structure, resulting in an increase in the type~I error rate. Similarly, the hierarchical Bayesian model assuming a single common cluster across all baskets (i.e., full pooling) successfully controlled the type~I error rate at the one-sided 5\% level when only one cluster was present; however, when heterogeneity in efficacy existed across clusters, this model also exhibited inflation of the type~I error rate. Regarding the 90\% credible interval widths, the proposed method yielded narrower intervals compared with the approach without clustering, indicating improved estimation efficiency. Several existing borrowing methods for basket trials, such as the EXNEX framework
\citep{Neuenschwander2016} and the calibrated hierarchical Bayesian (CHB)
model \citep{ChuYuan2018CHB}, address a related but somewhat different objective, namely robust borrowing under prespecified exchangeability structures. In contrast, the proposed framework focuses on learning data-driven borrowing groups from response trajectories when the latent grouping structure is unknown. For this reason, the primary simulation study emphasized comparators that either do not learn borrowing groups or learn them from simpler efficacy summaries. In addition, we considered an EXNEX comparator as a supplementary benchmark. In our simulation settings, the EXNEX comparator showed behavior closer to that of the No-cluster approach, suggesting comparatively limited borrowing under this specification.

A key limitation of the proposed approach is its tendency to over-partition baskets when the true underlying structure is homogeneous. This behavior was evident in Scenario~1 and may lead to unnecessary complexity in the borrowing structure. In addition, the current implementation treats response categories as nominal in the clustering distance and uses a pragmatic uniform regularization when no outgoing transitions are observed from a given state. These design choices improve simplicity and numerical stability, but they may not fully reflect the ordinal nature of tumor response or clinically realistic transition patterns. In practice, the final judgment will likely be made by comprehensively considering both the results within each cluster and the clinical differences among clusters. Future research should explore alternative or complementary clustering algorithms that are more robust in single-cluster scenarios within the framework of basket trials.

In conclusion, we proposed a trajectory-informed clustering framework for data-driven information borrowing in basket trials. Obtaining transition probability matrices in basket trials enables dynamic characterization of treatment response pathways (e.g., SD $\rightarrow$ PR $\rightarrow$ CR $\rightarrow$ PD) across tumor types, which may help inform the design and planning of later-phase trials~\cite{beyer2020multistate, krishnan2021multistate}. When latent heterogeneity was present in the considered simulation settings, the proposed method more often recovered the intended basket groupings than ORR-only clustering and yielded favorable borrowing performance relative to simpler alternatives. However, the homogeneous scenario also showed that the current silhouette-based implementation can over-partition baskets when no latent heterogeneity is present.

\section*{Data availability statement}
No new data were created or analyzed in this study. Data sharing is not applicable to this article.

\section*{Ethics statement}
This study did not involve human participants or animal experiments requiring ethical approval.

\section*{Conflict of interest statement}
The authors declare no conflict of interest.

\section*{Funding statement}
This research received no specific grant from any funding agency in the public, commercial, or not-for-profit sectors.

\newpage
\bibliography{main.bib}
\bibliographystyle{unsrtnat}



\end{document}

%% file: def.tex
\def\bpi{{\text{\boldmath $\pi$}}}

\def\bpih{{\widehat \bpi}}

\def\bpio{{\overline \bpi}}

\def\y{{\text{\boldmath $y$}}}

\def\N{{\text{\boldmath $N$}}}

\def\P{{\text{\boldmath $P$}}}

\def\Nh{{\hat \N}}
\def\Ph{{\hat P}}
\def\bPh{{\hat \P}}

\def\wh{{\hat w}}